\documentclass[
superscriptaddress,
preprint,
bibnotes,
 amsmath,amssymb,
 aps,
 pra,
]{revtex4-1}
\usepackage{graphicx}
\usepackage{epstopdf}
\usepackage{float}
\usepackage{rotating}
\usepackage{array}

\begin{document}

\title{$\hbar$ as a Physical Constant of Classical Optics and Electrodynamics}

\author{R\'{e}al Tremblay}
\affiliation{Centre d'optique, photonique et laser (COPL), D\'{e}partement de physique, g\'{e}nie physique et optique, Universit\'{e} Laval, Qu\'{e}bec, G1V 0A6, Canada}
\author{Nicolas Doyon}
\affiliation{D\'{e}partement de math\'{e}matiques et de statistiques, Universit\'{e} Laval, Qu\'{e}bec, G1V 0A6, Canada}
\author{Claudine N\`{\i}. Allen}
\affiliation{Centre d'optique, photonique et laser (COPL), D\'{e}partement de physique, g\'{e}nie physique et optique, Universit\'{e} Laval, Qu\'{e}bec, G1V 0A6, Canada}

\date{\today}

\begin{abstract}
The Planck constant ($\hbar$) plays a pivotal role in quantum physics.  Historically, it has been proposed as postulate, part of a genius empirical relationship  $E=\hbar \omega$ in order to explain the intensity spectrum of the blackbody radiation for which classical electrodynamic theory led to an unacceptable prediction: The ultraviolet catastrophe. While the usefulness of the Planck constant in various fields of physics is undisputed, its derivation (or lack of) remains unsatisfactory from a fundamental point of view. In this paper, the analysis of the blackbody problem is performed with a series expansion of the electromagnetic field in terms of TE, TM modes in a metallic cavity with small losses, that leads to developing the electromagnetic fields in a \textit{complete set of orthonormal functions}. This expansion, based on  coupled power theory, maintains both space and time together enabling modeling of the blackbody's evolution toward equilibrium. Reaching equilibrium with a multimodal waveguide analysis brings into consideration the coupling between modes in addition to absorption and emission of radiation. The properties of the modes, such as spectral broadening, losses and lifetime, then progressively become independent of frequency and explains how equilibrium is allowed in good conductor metallic cavities. Based on the free electron relaxation time in gold, a value of $\hbar = 1.02 \times 10^{-34}$ J$\cdot$s for the reduced Planck constant is found and the uncertainty principle is also emerging from this \textit{a priori} classical study. The Planck constant is then obtained no longer as an ad hoc addition but as a natural consequence of the analysis taking boundary conditions into account as into optical resonators. That analysis based on finite-spacetime paradigm, also shine new light on the notion of decoherence in classical optics and electrodynamics. 
\end{abstract}

\maketitle

\section{Introduction}

The ninetieth century saw great progress in both electromagnetism and thermodynamics with radiative heat transfer studies positioned between both. The latter studies of light-matter interactions led Gustav Kirchhoff to introduce the concept of perfect black bodies and the universality of their spectral radiance as a function of wavelength and temperature~\cite{Kirchhoff60_Blackbody}. As light was identified with electromagnetic (EM) waves, the blackbody became an ideal tool to connect energy notions with the thermodynamics of electromagnetic radiation. The initial EM analysis of thermal blackbody radiation by Lord Rayleigh and J. Jeans is well known to have supported the introduction of a light quantum due to a discrepancy in the radiative energy density distribution at high frequencies, dubbed ultraviolet catastrophe. Great successes of the radiation quanta in explaining both the photoelectric effect and blackbody radiation led the EM field theory and the Maxwell's equations to fade away from the spotlight. Over the years, semi-classical models without light quantization turned out to suffice in many cases of light-matter interaction, including the photoelectric effect~\cite{Lamb95_Anti-photon, Beck27_ClassicPhotoelectric}. Besides, time-independent Schr\"{o}dinger and Maxwell wave equations can both yield Helmholtz equations or other mathematically similar forms with a correspondence between potential and index of refraction. The reader can refer to~\cite{MHan_StoryofLight} for a review on the wave-particle notions. Let us present a citation from the author (p.9), about the Klein Gordon equation of the relativistic quantum mechanics; "An equation for a zero-mass particle of spin one (photon) in relativistic quantum mechanics turn out to be none other than the classical wave equation for the electromagnetic field of the 19\textsuperscript{th} century that predates both relativity and quantum physics". Being inherently relativistic and adaptable to  Hamiltonian formulation, the EM field theory was revisited with quantum electrodynamics (QED), a theory which explains with excellent precision experiments, as example, Lamb displacement, hydrogen hyperfine structure, and serves to analyze with success many others interactions. QED still however calls upon infinite plane wave approximations. An infinite plane wave basis used for the spatial dependence of the EM field is in contradiction with the finite size of a blackbody cavity, which incidentally should be seen as an optical resonator with boundary conditions allowing a penetration depth of the EM field in the walls. According to classical physics, the blackbody problem is analyzed under the hypothesis that  atoms located on  the cavity wall can only absorb and/or emit electromagnetic energy thus giving rise at thermodynamic equilibrium to a system of stationnary waves (stationnary modes) in the cavity. Studying perfectly reflecting walls and a system at equilibrium lead to a reduction of the electomagnetic modal structure from $\vec{E}(\vec{r},t)$ to $g(t)\vec{E}(\vec{r}).$ It is thus not surprising that classical electromagnetism fails to explain how equilibrium is reached since it defines equilibrium as a time and a space separated functions and neglects a very important phenomenon of multimodal propagation in a physical medium, namely: mode coupling. What really leads to the ultraviolet catastrophe in the analysis of this problem is thus the fact that the finite aspect of space-time only appears in the evaluation of the number of allowed modes in the cavity by supposing that they are associated to $TEM(\nu)$ waves with multiple perfect reflections along a given direction in the cavity.  In fact, in order to correctly solve this problem, on must, along the chosen direction, use pairs of $TEM$ waves of direction $\pm \theta_n$ fulfilling limit condition (transversal resonance) with respect to the surface.  This leads to developing the electromagnetic field in TE-TM modes, a set of \textit{complete and orthonormal functions} for the quantization of a field in a box.

By applying the theory of guided light propagation to the blackbody-resonator problem, we could introduce solutions in terms of finite field modes and also gain valuable insight into the dynamical processes allowing a blackbody to reach thermal equilibrium.The solution in finite spacetime with eigenvalues and eigenfunctions leads to the normalization of the exchange of energy between electrons in the volume $S\delta$ ($S$ being the cavity area and $\delta$ being the penetration depts of the EM field) and the EM modes oscillating in the total volume ($S\delta+V$). The Planck constant ($\hbar$) then appears as a normalization factor with respect to the characteristics of gold (as exemple). To the knowledge of the authors it is the first time an analysis of the Blackbody radiation including "power mode coupling" is presented. This mechanism, understandingly overlooked in Planck's papers, also shine new light to the notion of decoherence in classical electrodynamics. The reader will find in table~\ref{Table 1} a comparison between the approach proposed in this paper and the one generally presented in quantum physics.
\newpage
\begin{table}
    \includegraphics[width=15cm]{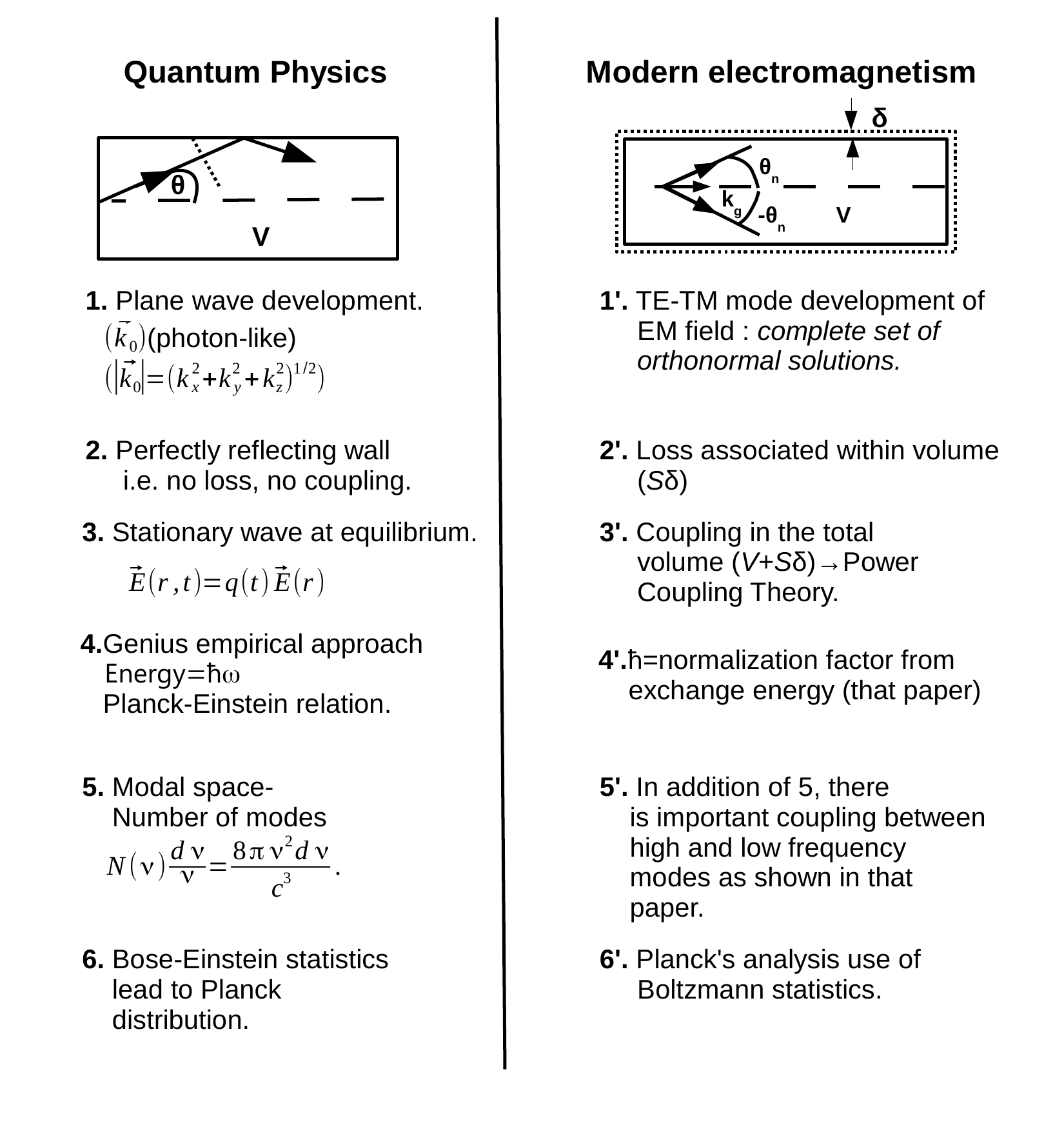}
    \caption{Comparison between quantum physics analysis and modern EM analysis of Blackbody radiation.}
    \label{Table 1}
\end{table}

An analysis to be cited~\cite{THBoyer_BlackbodyRadiationSpectrum,THBoyer_BlackbodyRadiationSpectrumANDSpacetime} based on zero-point radiation, pre-"suppose" the existence, at zero temperature, of "classical homogeneous fluctuating radiation", and as, now, in quantum physics with "quantum vacuum fluctuations", use an "ad hoc" correction to the definition of the equilibrium.

In this article, we will first present a few reminders of relevant waveguiding topics: the expansion of an EM field into ideal fundamental modes, the EM fields in resonant monomode metallic cavities with a modified analysis conserving a privileged propagation direction, and EM fields in multimode metallic cavities. This will introduce mode coupling as an essential feature of EM multimode guided propagation allowing fast power exchange between the modes. Second, we will turn to the study of blackbody resonators in the multimode regime where mode coupling, in addition to absorption and emission of radiation, is needed to reach thermal equilibrium. The average EM energy per mode in the cavity will be evaluated knowing that when low-loss multimode guided propagation is in equilibrium, all modes possess the same attenuation coefficient. To relate the EM energy to the cavity temperature, optimal energy exchange between the modes and the cavity walls will be examined at maximum synchronization with the free electron mean relaxation frequency $\omega_\tau$, related to their mean free path between collisions. The reduced Planck constant $(\hbar)$ and the radiative energy density distribution will be shown to emerge naturally from this relation obtained from EM analysis, $\hbar$, becoming a derivable physical constant. The uncertainty principle is also emerging from this \textit{a priori} classical theory. Furthermore, from a microscopic-macroscopic analysis, a frequency relation between the gold (reference) and others materials of a large group of good metallic conductors will be derived, thus establishing a link between the microscopic, $\omega_\tau$, matter oscillator (even magnetic one), and the macroscopic Planck distribution aspect of that problem.

\section{Electromagnetic field representation in fundamental normal modes}
\label{sec:EMF Rep}

\begin{figure}
    \includegraphics[width=10cm]{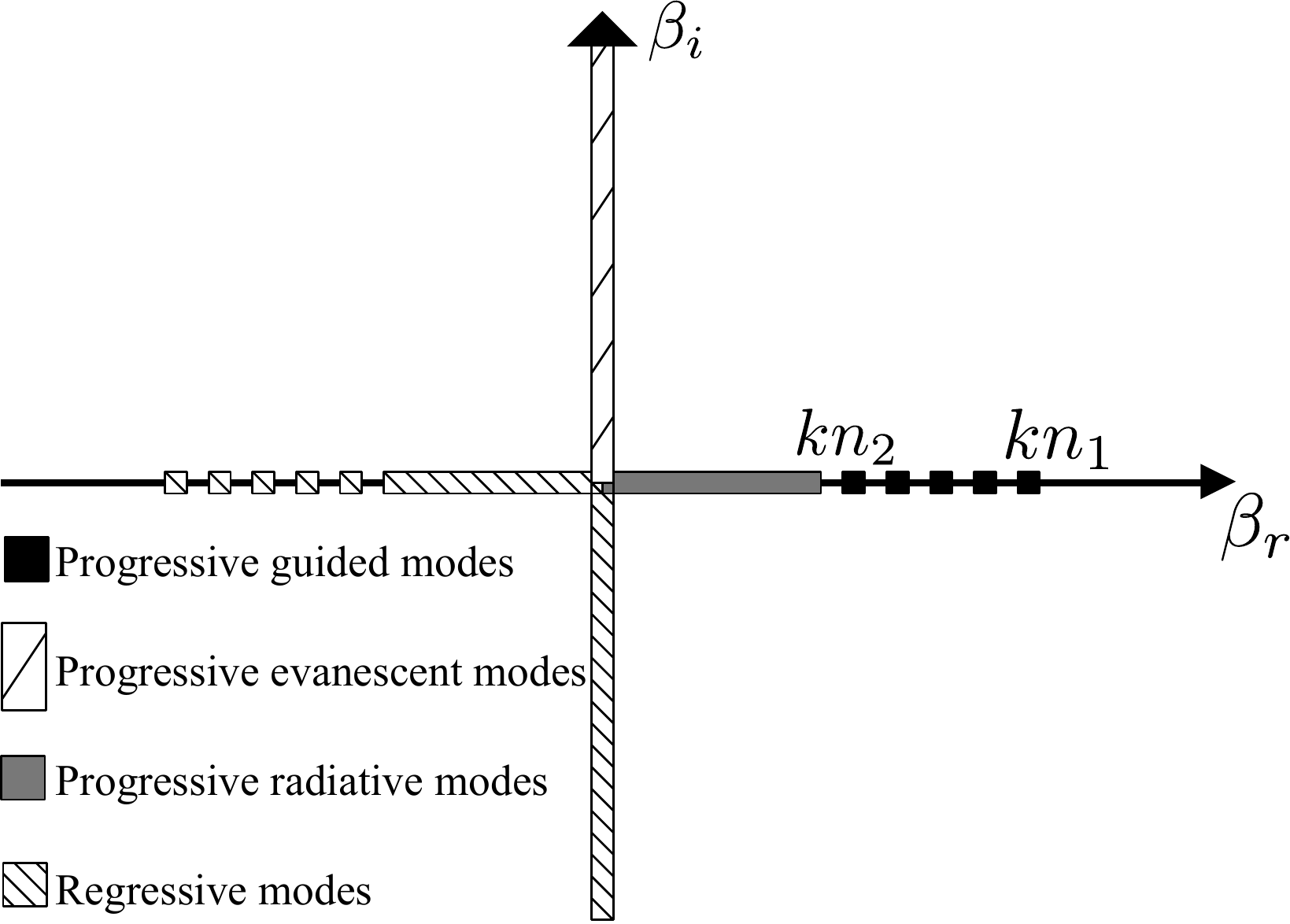}
    \caption{Modal domains of a waveguide with refractive index $n_1$ in a surrounding medium with refractive index $n_2 < n_1$, where $\beta_r$ and $\beta_i$ are the real and imaginary part of the modal propagation constant respectively.}
    \label{Figure1}
\end{figure}

Guided, radiative and evanescent modes of waveguides constitute a complete and orthonormal set~\cite{Kogelnik79_TheoDielectricWaveguides}. Such a set can therefore form a basis for a series expansion of an EM field. Figure~\ref{Figure1} represents the domains of these modes for a waveguide of refractive index $n_{1}$ in a medium of refractive index $n_{2}<n_{1}$, where $\beta$ is the propagation constant of the observed mode.Without coupling, the power in each guided and radiative mode is transported individually since $\beta$ is real. In the case of evanescent modes having pure imaginary $\beta$, a combination of progressive and retrograde evanescent modes is necessary to transport energy, they cannot do it individually ~\cite{Kogelnik79_TheoDielectricWaveguides, Marcuse91_TheoDielectricWaveguides}.

\newpage
 One can expand the EM field of an irregular waveguide in terms of ``ideal (normal) modes''. The terms of the expansion, then function of the propagation direction $z$, are solutions to a set of coupled differential equations. Another method exists to obtain a set of coupled differential equations describing  mode coupling. One can formulate the EM field of the irregular waveguide at a point $z_o$ in terms of the modes of an hypothetical local waveguide. These modes are called ``local modes''. However in this article, the first description based on ideal (normal) modes will be used as it is more appropriate for regular waveguides with ideal geometry and therefore lends itself well to the analysis of the blackbody problem. For a more detailed analysis of ideal mode coupling, the reader may refer to the third chapter of the book by D.~Marcuse~\cite{Marcuse91_TheoDielectricWaveguides} which we will use in section~\ref{subsec:MGPMC} of this paper. 

It is important to note that boundary conditions are not applied explicitly to obtain the solution that will be described. The ideal modes expansion method generates solutions that apply to all space, since the boundary conditions are contained in the mode equations composing the expansion. These boundary conditions are automatically verified by the solutions, thus there is no need to bring these relations later in the process.  This modal expansion also speeds up numerical simulations in general by one or two orders of magnitude due to the finite number of modes comparatively to the spatiotemporal model which deals with base equation of infinite dimensionality.

\section{Resonant metallic cavities}
\label{sec:RMetallicC}
\subsection{Monomode metallic cavities in modern electromagnetism}

At first, we will follow an analysis that can be found in electromagnetism textbooks, see for example J.~D.~Jackson~\cite{Jackson98_ClassicalED}, and we will later adapt it to the type of analysis done in quantum physics. The EM field inside the metallic cavity will be expanded into ideal (normal) stationary modes, \textit{i.e.} normal modes from short-circuited guided propagation along the $z$ axis which allows us to maintain a preferential orientation. We start by choosing a metallic cavity of rectangular shape, in order to reproduce the case of "quantization in a box", with dimensions $a$, $b$ and $d$, where $d$ is the depth of the cavity along the $z$ axis.

\begin{figure}[H]
	\begin{center}
	\includegraphics[width=10cm]{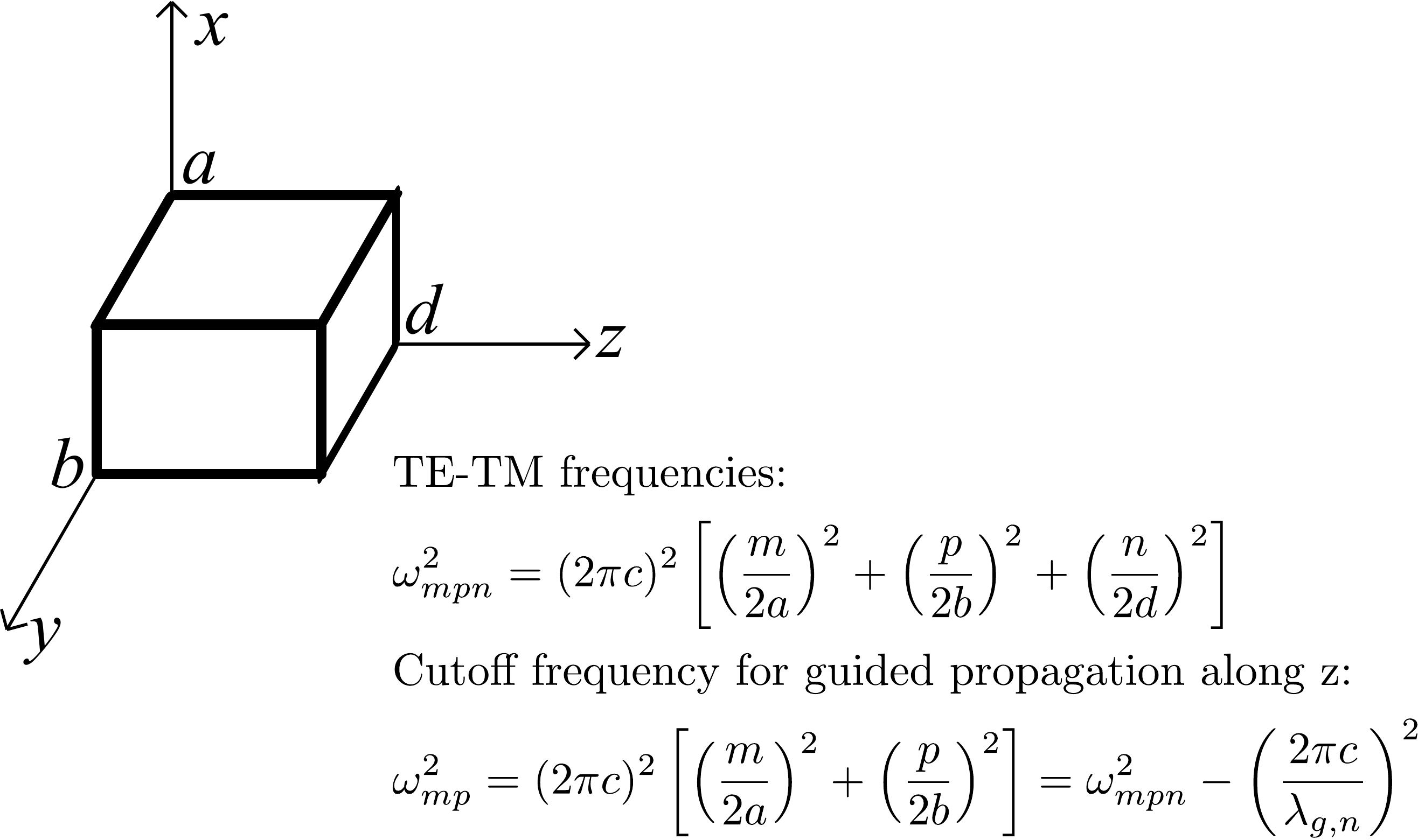}
	\caption{Geometrical parameters of a rectangular metallic cavity and associated frequency relations. Since the $z$ axis is designated as the axis of propagation and observation, $\lambda_{g,n}=2d/n$.}
	\label{Figure2}
	\end{center}
\end{figure}

For our modal expansion, we will label the mode in the $x$ direction with $m$, in the $y$ direction with $p$ and in the $z$ direction with $n$. Considering power losses in the cavity, one finds that the energy stored inside the cavity for a mode $(mpn)$ is given by the following equation:
\begin{equation}
	\label{eq:StoredEnergy}
	U_n = \frac{\omega_{mpn}^2 d}{\pi c^2} F_{trans}
\end{equation}
where
\begin{equation}
	\label{eq:Ftrans}
	F_{trans} = \begin{Bmatrix} \epsilon\\ \mu \end{Bmatrix} \frac{\pi^2 c^2}{4 \omega_{mp}^2} \int_A{|\psi(x,y)| da} \;\; \begin{Bmatrix} TM\\ TE \end{Bmatrix}
\end{equation}
with $\psi (x,y)$ representing the distribution of the transverse field associated to a $(mpn)$ mode where $n=2d/\lambda_{g,n}$.

At this step, authors often express the cavity volume $V$ explicitly in order to determine the cavity's $Q$-factor:
\begin{equation}
	\label{eq:Qfactor}
	Q = \frac{\mu}{\mu_{conductor}} \left(\frac{V}{S\delta}\right) \times \left(\textup{Geometrical Factor}\right)
\end{equation}

where $V$ and $S$ represent respectively the volume of the cavity and its surface area while $\delta$ is the penetration depth of the field inside the walls of the cavity. The relationship (\ref{eq:Ftrans}) is also used to determine the well-known density of modes
\begin{equation}
N(\nu)d\nu/V=8\pi \nu^2 d\nu/c^2
\end{equation}
where $N(\nu)d\nu$ represents the number of modes in the frequency interval $[\nu,\nu+\delta\nu]$. Usually, one would obtain an energy density by solving an integral over only one frequency band which is also equivalent to using the cavity volume $V$ explicitly.

However, we will not make $V$ explicit at the moment to keep the "behavior of time and space coupled". We will continue our analysis with the short-circuited propagating wave approach, which maintains the analogy with the one-dimensional harmonic oscillator by preserving a preferential direction of propagation. Since $d=n\lambda_{g,n}/2$, one can rewrite Eq.~\eqref{eq:StoredEnergy} in the form:
\begin{equation}
	\label{eq:QuantaStoredEnergy}
	U_n = \frac{n\lambda_{g,n} \omega_{mpn}^2}{2 \pi c^2} \: F_{trans} = \frac{n \omega_{mpn}}{v_{gr,n}} \: F_{trans}
\end{equation}
where $v_{gr,n}$ is the group velocity of the guided short-circuited mode $n$ in an empty cavity as defined by
\begin{equation}
	\label{eq:GroupVelocity}
	v_{gr,n} = c \sqrt{1 - \frac{\omega_{mp}^2}{\omega_{mpn}^2}}
\end{equation}
and $F_{trans}$ is defined in Eq.~\eqref{eq:Ftrans}.

Important results are obtained by analyzing the field directionally along $z$ for set $m$ and $p$ values. This step is essential to establish a link with quantum mechanics as the observable of the stored energy in the cavity can then be expressed as:
\begin{equation}
	\label{eq:QMStoredEnergy}
	U_{n} = \frac{n \omega_{mpn}}{v_{gr,n}} \: \left[ \chi_{mp}\right]_{TM,TE} \rightarrow \frac{n \nu_{mpn}}{c} \: \left[ \chi_{mp}\right]_{TM,TE}
\end{equation}
where the rightmost expression is derived for the asymptotic limit far from the cutoff frequency $(\omega_{mpn} \gg \omega_{mp})$ with $\chi_{mp}$ designating a constant for TM and TE modes.

If we assume lossless, perfectly reflective cavity walls, a definite link is thus established between modern electromagnetism and quantum physics through the quantization $n\nu_{mpn}$ of the energy. However, the effects of losses associated to finite conductivity are already included in $\omega_{mpn}$ which is linked to the resonance frequency $\omega_{0(mpn)}$ of an ideal lossless cavity. Large values of ohmic losses are always causing a decrease in the resonance frequency, as seen in the following relation:
\begin{equation}
	\label{eq:CavityLosses}
	\omega-\omega_0 = \Delta\omega \approx -\frac{\omega_0}{2Q}.
\end{equation}
Thus our analysis does not make the typical, yet unrealistic, assumption of perfectly conducting walls. Let us add that when the system is left undisturbed, the energy initially stored in the mode $n$ of a realistic cavity will decrease exponentially with a time constant $\tau$ inversely proportional to $Q$, \textit{i.e.} $\tau \: \propto \: \left(\omega_0 /Q \right)$. The outward power flow is then equal to:
\begin{equation}
	\label{eq:PflowOut}
	P_n = v_{gr,n} U_n = n\omega_{mpn} \: F_{trans}.
\end{equation}
In the classical electromagnetic approach to the blackbody problem, the number of modes in the modal space was determined in conditions more restrictive than the asymptotic limit of the group velocity of all modes in the cavity tending toward $c$. A TEM wave method was used with $v_{gr}$ taken exactly equal to $c$, but the propagation of infinite TEM waves is forbidden \textit{individually} inside a hollow metallic waveguide. How can one easily apply the finite boundary conditions of a blackbody in this situation while preserving the coupling between space and time variables? It is perhaps not surprising that these approximations led to the failure of classical electrodynamics for the blackbody problem. Interestingly in quantum physics, one implicitly makes the $v_{gr}=c$ assumption of classical electrodynamics when quantizing the energy according to $E=nh\nu$ ($h\nu$ and the corresponding states $|\psi_n\rangle$).

In contrast, the blackbody boundary conditions are naturally satisfied by the TE, TM normal mode description from short-circuited guided propagation. A few observations readily follow from this description: 1) the bounded TE, TM modes link local phenomena to their environment inducing a related decoherence; 2) a wave packet formulation can be based on the group velocity $v_{gr}$ and the losses from the cavity environment; 3) the need for a renormalization process is eliminated since the TE, TM modes inherently satisfy special relativity, and are a \textit{set of complete orthonormal solutions}; 4) this TE, TM modes picture is valid for arbitrarily shaped cavities as long as their size is sufficiently greater than the wavelengths involved. Because of the latter, one can neglect a slow envelope modulation of the EM field in space as well as small perturbations near the cavity walls in the asymptotic limit $v_{gr} \rightarrow c$. The spatially rapidly varying part of the EM field inside an arbitrarily shaped cavity then corresponds to the TE and TM modes of a rectangular cavity; each 3-dimensional period of this spatial variation will be henceforth referred to as TE-TM cell. In this case, all cavities have the same relation for the electromagnetic mode density which is also the one found in classical electromagnetism.

\subsection {Multimode guided propagation and mode coupling}
\label{subsec:MGPMC}

As was noted by Planck himself~\cite{Kuhn78_BlackbodyHistory}, a complete description of blackbody radiation requires another mechanism than light emission and absorption to explain how energy is redistributed between different frequencies to reach the thermal equilibrium distribution from an initial arbitrary distribution. In the present paper, we  show that the mode coupling mechanism introduced from guided EM theory in section~\ref{sec:EMF Rep} has the necessary characteristics to ensure such an energy redistribution. We thus return to the EM field representation in ideal (normal) modes in the case of multimodal propagation with mode coupling. This involves an infinite set of integro-differential equations. However, this is an unsolvable problem except if certain conditions can be set to reduce its complexity. Marcuse demonstrated that it is possible to obtain a set of coupled equations bringing into play the average power in each mode instead of their amplitude~\cite{Marcuse91_TheoDielectricWaveguides}. This demonstration holds for waveguides of arbitrary length, as long as the coupling is weak. It is based on the experimental observation of the average power of a mode in highly multimodal propagation with weak coupling. After an adaptation length, this power averaged in $z$ for a given mode $\nu$ is of the form:
\begin{equation}
	\label{eq:Pmoy}
	P_{\nu}^{(moy)}(z) = A_{\nu}\exp(-\sigma z)
\end{equation}
where $\sigma$ is independent of $\nu$ and will be found to represent the eigenvalues of a set of homogeneous equations.

In the absence of coupling for multimodal guided propagation, an attenuation coefficient $2\alpha_{\nu}$ can be attributed to each uncoupled mode to represent the power flow carried by a mode along the waveguide as follows:
\begin{equation}
	\label{eq:Puncoupled}
	P_{\nu}(z) = P_{\nu}(0)\exp(-2\alpha_{\nu} z)
\end{equation}
where $P_{\nu}(0)$ corresponds to the power flow of mode $\nu$ at $z=0$. The shape of the power distribution is stable only when the mode, or a small subset of modes, with smallest losses carries power while the other modes have disappeared completely. The total power carried by the set of all modes does not decrease exponentially in this case.

\begin{figure}
	\centering
	\includegraphics[width=10cm]{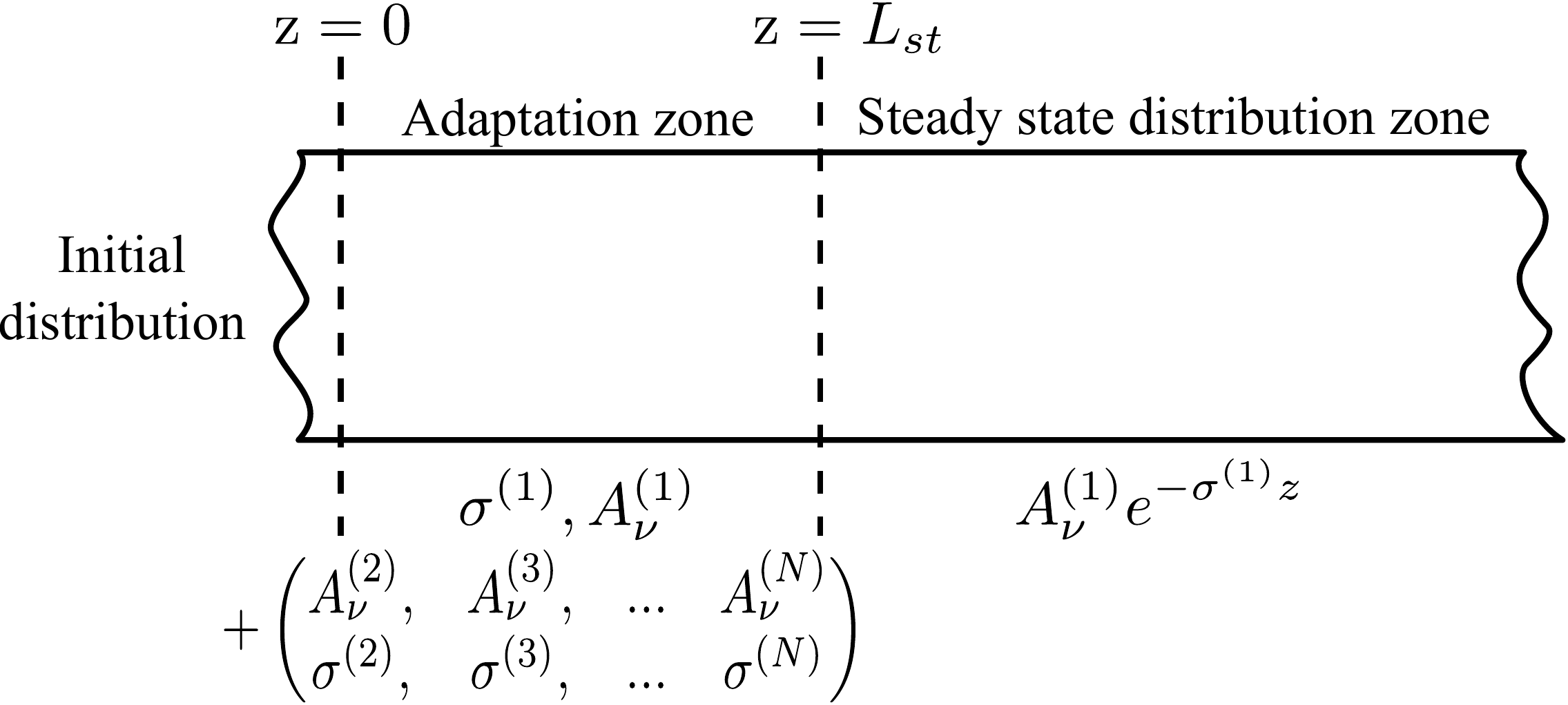}
	\caption{The initial power distribution at $z=0$ enters an adaptation zone with a constructive chaotic power distribution up to the beginning of the steady state power distribution at $z=L_{st}=\frac{\ln K}{\sigma^{(2)}-\sigma^{(1)}}$ where $K$ is related to the tolerated signal-to-noise ratio. No physical meaning is associated to the eigenvalues $\sigma^{(N)}$ and eigenvectors $A^{(N)}$ of order $N$ superior to 1.}
	\label{Figure3}
\end{figure}

In the presence of coupling induced by a mechanism independent of $z$, the power redistributes itself between the different modes, following the combined effects of individual losses and coupling exchanges. If this coupling is weak enough to keep losses very low, but strong enough to overcome individual losses, it tends to equalize the power carried by each mode and to establish a stable distribution of its power characterized by Eq.~\eqref{eq:Pmoy}. The ratio of average power carried by each mode in relation to a reference mode then no longer depends on the length coordinate. The same attenuation coefficient can therefore be attributed to all modes.The weak mode coupling regime allowed Marcuse to obtain, using the perturbation theory, the following average power coupling relation:
\begin{equation}
	\label{eq:Pdiffeq}
	\frac{dP_{\mu}^{(moy)}}{dz} = -2\alpha_{\mu}P_{\mu}^{(moy)} + \sum_{\nu=1}^N C_{\mu\nu}\left( P_{\nu}^{(moy)} - 		 P_{\mu}^{(moy)}\right)
\end{equation}
where $\mu$ and $\nu$ indicate modes in this set and $C_{\mu\nu}$ represents the coupling coefficient as noted by Marcuse. If at a location $z$, only the mode $\mu$ carries energy, (\ref{eq:Pdiffeq}) reduces to a simple and intuitive representation 
$$
\frac{d\vec{P}_{\mu}}{dz}=-\left(2\alpha_{\mu}+\sum_{\nu=1}^{N} C_{\mu\nu}\right)\vec{P}_{\mu}
$$
where the term of thermic loss is augmented (or compounded) by the sum of all the coupling coefficients indicating that the mode $\mu$ loses power to all the other modes. In the complementary case of a mode $\mu$ at location $z$ not carrying any power, we then obtain from (\ref{eq:Pdiffeq}) 
$$
\frac{d\vec{P}}{dz}=\sum_{\nu=1}^{N}C_{\mu\nu}\vec{P}_{\nu}.
$$
The positive derivative shows that  the mode $\mu$ gains in power at the detriment of all the other modes. A set of homogeneous equations with $N$ unknown is obtained from~\eqref{eq:Pmoy} and~\eqref{eq:Pdiffeq}. The $N$ eigenvectors $A_{\nu}^{(n)}$ and associated eigenvalues $\sigma^{(n)}$ result from the solutions to this set of equations \textit{i.e.}. The evolution of the propagating field is represented in figure~\ref{Figure3}.  
 
 The steady state power distribution obtained by Marcuse decreases exponentially with $\sigma^{(1)}$ being independent of $\nu$:

\begin{equation}
	\label{eq:SSpower}
	P_{\nu}^{(moy)}(z) = \left[ \sum_{\nu=1}^N A_{\nu}^{(1)} P_{\nu}(0) \right] A_{\nu}^{(1)} \exp( -\sigma^{(1)}z)
\end{equation}

for $z\geq L_{st}$ with the parameters represented in figure~\ref{Figure3}. $L_{st}$ defines the distance at which this steady state is reached with

\begin{equation}
	\label{eq:SSlength}
	L_{st} = \frac{\ln{\kappa}}{( \sigma^{(2)} - \sigma^{(1)} )}
\end{equation}

where the value of $\kappa$ is adjusted according to experimental observations. The relative notion of stability (equilibrium) of information propagation expressed by the tolerance to a given level of noise is represented by relation (\ref{eq:SSlength}) where $\kappa$ stands for the tolerated noise level. This relation shows that the equilibrium state is nondeterministic which is related to the fact that uncertainty relationships   can  be obtained  from classical electromagnetism in a finite-spacetime as we will show in section~\ref{subsec:Uncert}. After the adaptation zone $0\leq z\leq L_{st}$, the distribution of power versus mode number depends only on the shape of the first eigenvector $A_{\nu}^{(1)}$ and is completely independent of the initial power distribution. Mode coupling thus tends to distribute the power evenly over all the modes as long as it does not introduce strong radiation losses. This condition is fully respected in the case of hollow metallic waveguides and also in low loss dielectric guides.

As can be expected for equilibrium systems, one cannot work its way backward from measures of the steady state to find out the initial power distribution over the modes (lack of reversibility). At best, Marcuse's model allows us to know which initial state $A_{\nu}^{(1)}$ should be used to reduce or eliminate the adaptation zone as noted by Marcuse.

\section{Multimode resonator and the blackbody problem}
\label{sec:Blackbody}

The typical approach to study the EM field in a closed cavity is to assume the system has reached its steady state in order to separate the time and space variables: $\vec{E}(\vec{r},t)=q(t)\vec{E}(\vec{r})$. However, by doing so we lose all the information about the transient state and  modal coupling effects, since the relation between space and time is eliminated. The coupled power equations theory summarized above can address this issue and also ensures that the blackbody is an ergodic system even from a purely electromagnetic perspective. Indeed, the stationary states of very high order in a closed cavity are preserved as can be expected for a low-loss multimodal system (highly reflecting walls) with relatively important mode coupling redistributing the energy between the modes. These stationary states are mostly established by stable propagation inside the cavity while the radiation within the adaptation zone contributes noise to the system.

Mode coupling was also confirmed experimentally~\cite{Marcuse81_OptFiber} through the observation of frequency independent losses as predicted by the exponential power decay~\eqref{eq:SSpower}. For all modes circulating in a realistic cavity, we can therefore define the same characteristic length $z_{\; 1/e}$ traveled over $\eta$ round-trips in a cavity of length $d$ before they have lost $1/e$ of their power:
\begin{equation}
	\label{eq:lifetime}
	z_{\; 1/e}=\left( \sigma^{(1)} \right)^{-1}=\eta d = v_{gr, n} \tau_{n}.
\end{equation}
The right hand side defines the corresponding lifetime $\tau_n$ of mode $n$ in the cavity using the group velocity $v_{gr, n}$ from Eq.~\eqref{eq:GroupVelocity}. We also note that the $v_{gr, \nu} \tau_{\nu}$ product is independent of frequency and must be the same for all modes $\nu$. When $\omega_{mp} \gg \omega_{mpn}$ with the limit $v_{gr, n} \rightarrow c$, the modal lifetime $\tau_n = \tau'$ also becomes frequency independent and the cavity's $Q$-factor from Eq.~\eqref{eq:Qfactor} will be:
\begin{equation}
	\label{eq:SpectralBroad}
	Q = \omega \tau' \approx \frac{\omega}{c\sigma^{(1)}}.
\end{equation}
Independence of frequency thus extends to spectral broadening of the modes as well, since $\Delta \omega = \omega /Q \approx c\sigma^{(1)}$. Once equilibrium is reached, all the individual modes have the same properties $\sigma^{(1)}, \eta, \Delta \omega$ and $\tau'$, except for $Q$ becoming a linear function of $\omega$.

\begin{figure}[H]
	\centering
    \includegraphics[width=10cm]{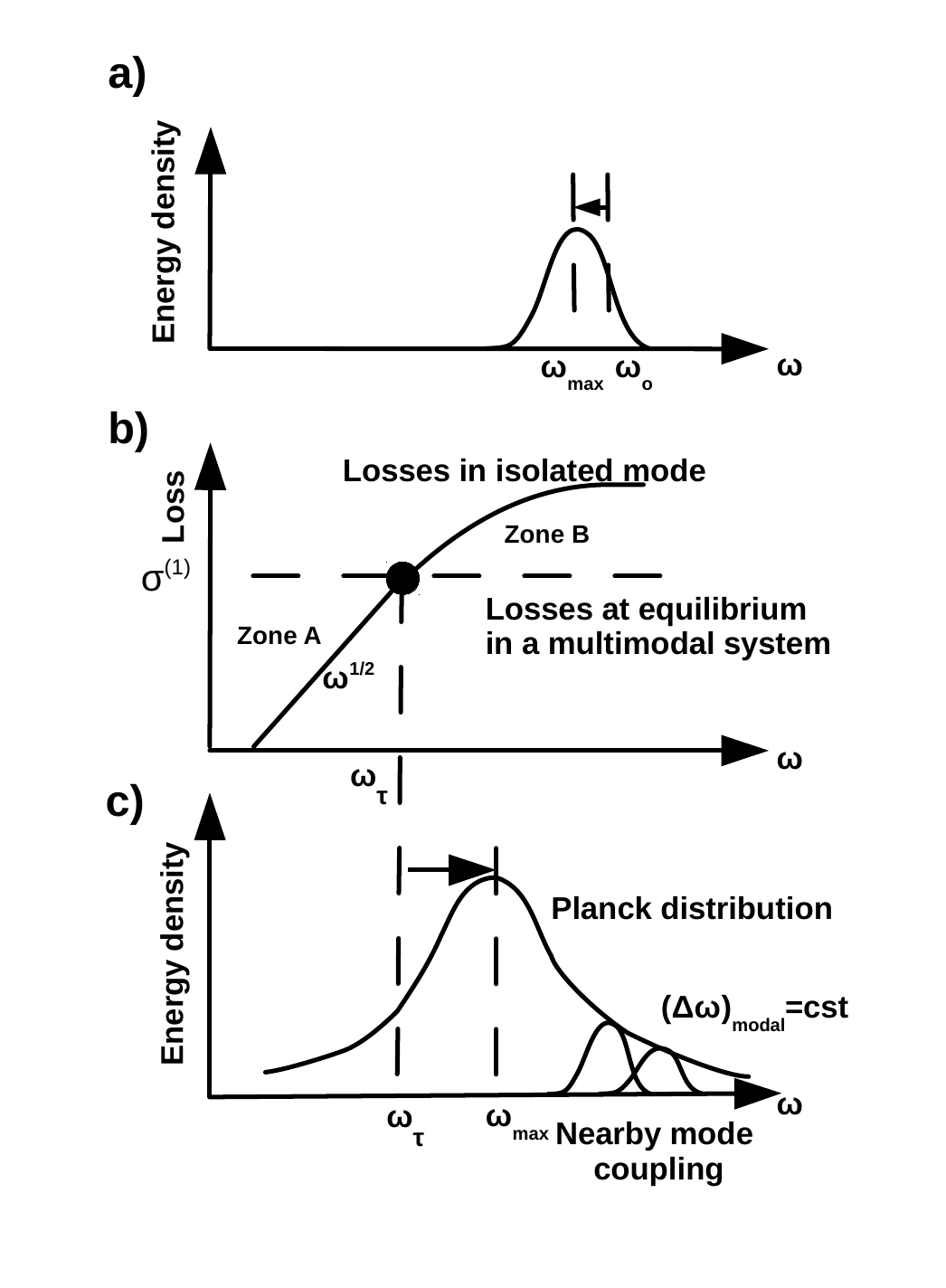}
    \caption{Schematic distribution for a) a monomodal system with $\omega_o > \omega_{max}$, b) losses of isolated modes (---) and loss $\sigma^{(1)}$ at equilibrium (- - -) for a multimodal system with representation of the two initial groups of modes in term of their resistance, c) resultant energy distribution at equilibrium where $\omega_{\tau}$ plays the role of $\omega_o$ and $\omega_{\tau} < \omega_{max}$.}
    \label{Figure4}
\end{figure}

To complete our analysis, the nature of the coupling between modes will be characterized. For the metallic blackbody cavity, ohmic losses are the dominating contribution to mode coupling. The Drude model tells us that absorption for each electromagnetic mode will be proportional to $\omega^{1/2}$ in the microwave frequencies while in the far-infrared region, it tends to a constant for $\omega \gg\omega_{p}$, (see figure~\ref{Figure4}.b)the plasma frequency. In the near infrared however, the absorption function is significantly dependent on $\omega_{\tau}$, which is a damping frequency linked to the free electron relaxation time~\cite{Muller-Kirsten04_EDIntroQuantum}. For an isolated mode with resonance frequency $\omega$ very far from the cutoff frequency $\omega_c$ --- The modes with $\omega$ close to $\omega_c$ do not obey this law, since their geometrical factor is different. Furthermore, the number of modes in the modal space no longer obeys the classical asymptotic relation. --- losses in the cavity are also proportional to $\omega^{1/2}$~\cite{Jackson98_ClassicalED}. It is more convenient to represent the losses for $\omega \gg \omega_c$ in function of $\omega$, $\omega_p$, and $\omega_{\tau}$ using the complex dielectric function $\epsilon_c$ as given by the Drude model~\cite{Ordal83_OptPropMetalsIR,Ordal85_OptProp14MetalsIR}:
\begin{equation}
\label{eq:DielectricFct}
\epsilon_c = \epsilon_1 +i \epsilon_2 = \epsilon_{\infty} - \frac{ \omega_p^{2} }{ \omega^{2} + i \omega \omega_{\tau} }
\end{equation}
where
\begin{eqnarray}
\omega_{p} &=& \frac{1}{2\pi c} \sqrt{ \frac{ 4\pi Ne^{2} }{m^*} }  \; \; \textup{cm}^{-1}, \label{eq:Frequency_1}\\
\omega_{\tau} &=& \frac{1}{2\pi c \tau}  \; \; \textup{cm}^{-1} \label{eq:Frequency_2}.
\end{eqnarray}
In these equations, $N$ is the free electron density, $e$ is the charge of the electron and $m^*$, it's effective mass. The high-frequency dielectric constant is represented by $\epsilon_{\infty}$ while the free electron relaxation lifetime is given by $\tau$. It should be noted that $\omega_{\tau}$ is evaluated from measurements of the complex dielectric function $\epsilon_c$.
 
Our analysis of the blackbody problem based on electromagnetism relies on the relation between $\omega_{\tau}$ and the temperature. Finite field volumes in the cavity (electromagnetic energy) and finite microscopic interaction volumes in the cavity walls (thermal energy) are exchanging energy by coupling. When the cavity sustaining the radiation field is kept at constant temperature, the best energy exchange will occur at the optimum synchronization frequency $\omega_{mpn}=\omega_{\tau}$. At this equilibrium, all modes travel the same number of round-trips and, in their asymptotic limit, have the same lifetime $\tau'$ in the cavity. We can then consider our resonant multimodal system with synchronization frequency $\omega_{\tau}$ equivalent to a lossless monomodal resonant system with the single resonant frequency $\omega_o = \omega_{\tau}$ (see figure~\ref{Figure4}.a). For the mode at $\omega_{\tau}$ and those nearby in frequency at a position $z$ where $P_{\mu}^{(moy)}$ is maximum, Eq.~\eqref{eq:Pdiffeq} yields:
\begin{equation}
\label{eq:Psteadystate}
\frac{dP_{\mu}^{(moy)}}{dz}= -\left[\sum_{\nu=1}^N C_{\mu\nu}\right]P_{\mu}^{(moy)}
\end{equation}
at thermal equilibrium. The exchange of thermal to electromagnetic energy and vice versa is due to the coupling term $\Upsilon \equiv\sum_{\nu=1}^N C_{\mu\nu}$. This energy exchange per unit of TE-TM cell in the propagation direction is expressed by equating Eq.~\eqref{eq:QMStoredEnergy} to the thermal energy as per the equipartition theorem:
\begin{equation}
	\label{eq:Equipartition}
	2\left(\frac{1}{2}kT\right) = \Upsilon \cdot \omega_{\tau}=\Upsilon \cdot\omega_{n}.
\end{equation}
The factor 2 comes from the transverse electric and magnetic fields contributing a degree of freedom each~\cite{Louisell73_QStatRad}. As there are $n$ cells in each independant propagation direction, the total energy in the mode of indice $n$ is:
\begin{equation}
E_n = n \;\omega_n \cdot \Upsilon.
\end{equation}
We can thus determine the value of the coupling constant $\Upsilon$ with a reliable value of $\omega_{\tau}$ at normal temperature ($293.15$ K). Since the definition of a Kelvin~\cite{Preston-Thomas90_IntTScale} (K) makes the temperature unit dependent on the property of a material, we must use a reference conductor satisfying the Drude model such as gold to ensure the accuracy of relation~\eqref{eq:Equipartition}. Br\"andli and Sievers have shown that measurements of $\epsilon_1$ and $\epsilon_2$ for gold in the infrared follow the Drude model very well, without using adaptive parameters for the number of free electrons per atom as is often done for other materials~\cite{Sievers72_FarIRSurfaceR}. Other arguments in favor of using the $\omega_{\tau}$ value of gold are its very good experimental agreement for the conductivity according to Ohm's law and negligible interband contributions to its absorption in the infrared. The average value measured for gold with different methods is $\left(\omega_{\tau}\right)_{gold}=2.11\times 10^2$~cm$^{-1}$~\cite{Ordal83_OptPropMetalsIR, Ordal85_OptProp14MetalsIR, AshcroftMermin76_SolidStatePhy}. According to~\eqref{eq:Equipartition}, we then obtain the constant value:
\begin{equation}
	\label{eq:hbar}
	\Upsilon_{gold} = 1.02 \times10^{-34} \; \textup{J}\cdot\textup{s} \cong \hbar.
\end{equation}

This result agrees well with the quantum mechanical analysis. Here, the Planck constant appears as a normalization factor between electromagnetic and thermal energy exchanges at thermal equilibrium. Reuniting these two domains, this result expands the signification of the Planck constant beyond the commutator function it fulfills in the quantization of a particle-electromagnetic field system. Electromagnetic coupling therefore plays an essential role: it enables the cavity to reach the absolute stable equilibrium required by the second law of thermodynamics. This equilibrium is achieved by minimizing cavity losses with power transfer from modes having high losses to modes having lower losses as well as with the well-known emission and absorption of radiation. To resume the "stationnary" states, equilibrium, is mostly established by stable propagation inside the cavity while the radiation within the adaptation zone contributes to noise of the system, as represented by the short-circuited propagation model of figure~\ref{Figure3}. 

Two different situations can occur in the process (figure~\ref{Figure4}). In the first case (4.a) where the losses are important, in a closed or quasi-closed system, it causes strong dissipative power transfert from high frequency modes (of higher loss) to lower frequency modes (of lower loss), and ultimately, it brings the collapse, at the equilibrium, in a few modes or only in one mode as in figure~\ref{Figure4}.a), where $\omega_o$ corresponds to the resonnant frequency of an equivalent system without loss, and where $\omega_o > \omega_{max}$. In the second case, for a multimodal system with very low loss and relatively strong coupling, as, for example, in metallic thermostatic cavities, the high frequency modes (higher loss), mainly, release power by intermodal coupling to low frequency modes (lower loss) and $\omega_{\tau}$, the relaxation frequency, of matter oscillators, plays an equivalent role as $\omega_o$, but now with $\omega_{\tau} < \omega_{max}$. Mode coupling thus tends to distribute the power evenly over all modes as long as it does not introduce strong radiation loss. That condition being respected in the cases of hollowo metallic resonator and in low loss dielectric resonators. It should be noted that in the first stage of the equilibrium process, modes of two frequency "zones" contribute to the coupling in different manners (see figure~\ref{Figure4}.b); in zone B ($\omega_n > \omega_{\tau}$) the resistivity of an equivalent system is initially positive, while in zone A ($\omega_n < \omega_{\tau}$) the resistivity is initially negative, meaning that its contributions modified the field without important contribution to $\rho_{op}$ (optical resistivity). In a way, the modes act, in part, as in a semi-conductor material and thus, our analysis presents an intuitive approach to the study of semi-conductor optical cavities following the finite-spacetime paradigm in classical electrodynamics. That research is out of the scope of that paper.\\

Also of importance, is the following remark about the "dissimilarity" of the modes related to finite-spacetime through their $\omega_c$, cutoff frequency, function of the mode order, and their distinct coupling effects. So it is not appropriate, in a finite multimodal system, to assert that "all elements are alike" and that "all dissipative effects" act to lower the maximum frequency of the resulting energy density spectral distribution. Modes of frequency $\omega_n < \omega_{\tau}$ act to enhance, by multimodal coupling, the $\omega_{max}$ of the spectral energy density distribution (Planck distribution). That effect is essential to the establishment of the equilibrium in the system. These two characteristics are essential to explain how the equilibrium is attainable within the finite-spacetime paradigm and the TE-TM field developpment based on a \textit{complete set of orthonormal functions}. Otherwise, as in the space-time paradigm where the photons are all alike, one must accept an "ad hoc" commutator function for $\hbar$, uncertainty relations as principles and "quantum vacuum fluctuations" as derived from $\hbar$ and uncertainty relations, that notion being used to explain the Casimir effect~\cite{PhysRevD.72.021301}.
\subsection{Planck's law}
The previous results obtained from a classical electromagnetic analysis demonstrate that the equilibrium state in the cavity is reached and maintained through $h\nu$ quanta characterizing exchanges of thermodynamic and electrodynamic energy. Each exchange of a quanta can be interpreted as a temporally finite excitation of a mode and in blackbody radiation conditions, the total number of excitations easily falls into the limit of large numbers. Since an electron can produce over time an arbitrary number of excitations fluctuating in energy, hence possibly but not always coupling to the same mode, the time-averaged energy distribution of these excitations will follow the distribution of energy density as derived by Planck~\cite{planck2012eight}, as example see page 19 of~\cite{eisberg1974quantum}. With the average energy of this distribution and the well-known density of modes $N(\nu)d\nu/V = 8\pi \nu^2 d\nu /c^3$, which can be calculated from either the modal approach followed here or the classical counting of standing waves in the cavity, we express Planck's law as follows:
\begin{equation}
	\label{eq:Planck}
	\rho(\nu) d\nu = \frac{8\pi h\nu^3}{c^3} \frac{1}{ e^{\nu / \nu_\tau(T)_{gold}} -1} d\nu \Leftrightarrow  \frac{\hbar \omega^3}{\pi^2 c^3} \frac{1}{ e^{\omega / \omega_\tau(T)_{gold}} -1} d\omega = \rho(\omega)d\omega.
\end{equation}
We chose to express the familiar relation for radiative energy density $\rho(\nu) d\nu$ with the damping frequency $\omega_\tau(T)$ instead of the usual $h/kT$ factor to emphasize the connection between temperature and materials. In the next section, we will analyze the relevance of other materials, than gold, by obtaining a parametric frequential relation linking all good conductors to Planck distribution.

\subsection{Good conductors metallic cavities}

A comparison against the gold must be established to obtain a radiative energy density distribution $\rho(\omega)d\omega$ obeying Planck's law as observed for several metals. This is done by introducing a comparison factor $G_X$ in the maximum synchronization relation~\eqref{eq:Equipartition} for an arbitrary metal labelled $X$, \textit{i.e.},

\begin{equation}
	\label{eq:CompFactor}
	G_XkT=Constant\cdot (\omega_{mpn})_X=Constant\cdot (\omega_{\tau})_X.
\end{equation}

The factor $G_X$ allows the scaling of temperature in other materials against the gold (as example) with $kT$ defined as $\hbar(\omega_{\tau})_{gold}$ we obtain $G_X=(\omega_{\tau})_X/(\omega_{\tau})_{gold}$ from the above relation. In the time domain, we can interpret this ratio as the relative number of cycles per unit of time over which an energy quanta is exchanged, which is consistent with the Planck constant units in joules per hertz. In the spatial domain, this corresponds to the ratio of TE-TM cells per unit of length over which coupling takes place, in relation with the weak losses and penetration depth of the electromagnetic field in the cavity walls. Additionally, different material will have a different density of electrons emitting and absorbing radiation, which will be included in $G_X$ via the ratio of free electron concentrations $N_e$. We must also include materials parameters influencing the overall system responds through the synchronization between the electromagnetic modal electronic and the damping frequencies.\\

The microscopic model of the Ohm's law implies~\citep{AshcroftMermin76_SolidStatePhy} that the courant-voltage relation relation at the input-output of a material is due to the fact that the associated electric field adds a small (regular) speed to the chaotic one of the free electrons in the material. Taking into account relation~\eqref{eq:Frequency_2}, one finds that~\citep{AshcroftMermin76_SolidStatePhy} the relation frequency is

\begin{eqnarray}
	\label{eq:RelaxFrequency}
	\omega_{\tau}(\textup{cm}^{-1})=\frac{N_e e^2 \rho}{m_e 2\pi c}
\end{eqnarray}

where \begin{itemize}
			\item[] $N_e =$ electron number/cm$^3$
			\item[] $e =$ electric charge 
			\item[] $\rho =$ resistivity
			\item[] $m_e =$ electron mass.
	  \end{itemize}

That relation~\eqref{eq:RelaxFrequency} applies only to materials corresponding to the Drude model, as for example, gold and silver. At normal temperature (293 K) their calculated results from~\eqref{eq:RelaxFrequency} are $(\omega_{\tau})_{Au}=2.11\times 10^{2}$cm$^{-1}$ and $(\omega_{tau})_{Ag}=1.42\times 10^{2}$cm$^{-1}$ while their mesured ones are $(\omega_{\tau})_{Au}=2.15 \times 10^{2}$cm$^{-1}$ and $(\omega_{\tau})_{Ag}=1.45 \times 10^{2}$cm$^{-1}$ from~\cite{Ordal83_OptPropMetalsIR,Ordal85_OptProp14MetalsIR}.\\

So, in the case of these materials, the parametric frequential relation $G_X$ is reduced to only ratios of electronic density and resistances; \textit{i.e.}

\begin{equation}
	\label{eq:FactorGold}
	G_X=\frac{(\omega_{\tau})_X}{(\omega_{\tau})_{gold}}=\frac{(N_e)_X}{(N_e)_{gold}}\cdot\frac{\rho_X}{\rho_{gold}}.
\end{equation}

The number of the materials obeying to relation~\eqref{eq:FactorGold} is quite small, $Al$ already departing of it by almost 13\%. The authors of reference~\cite{Ordal83_OptPropMetalsIR,Ordal85_OptProp14MetalsIR} note, without surprise, that at high frequencies, $Au, Ag$ and $Al$, follow well the free electron model, but that the overall picture does change for the others materials (interband and surface effects). Furthermore, in a metallic thermostatic cavities, at the microscopic level, the resulting resistivity is mainly determined by the modal high frequency phenomena, those modes releasing most of their power by coupling to the lower frequency modes (see figure~\ref{Figure4}). The dominant physical mechanism in our approach is ultra fast coupling compared to slower absorption and re-emission. Hence, one have to choose appropriate effective $N_e, m_e, \rho$. In classical electrodynamics, a "good conductor can be modelled as a perfect conductor with the idealized surface courant replaced by an equivalent surface courant, which is actually distributed throughout a very small, but finite thickness at the surface. Thus, the surface resistance has to replace the volume resistance; \textit{i.e.} $1/\sigma\rightarrow 1/\sigma\delta$, where $\sigma$ is the conductivity, and $\delta$ the skin depth"~\cite[chap.8]{Jackson98_ClassicalED}. In a preceding section~\ref{sec:Blackbody}, it was shown that the "modal" quality factor, in multimodal cavity resonator, was,

\begin{equation}
	\label{eq:MultimodalCavityResonator}
	(Q)_X=\frac{\omega_n}{c\sigma^{(1)}_X},
\end{equation}

where, here, the subscript X is for a chosen material. Remembering that, at equilibrium, "$\sigma^{(1)}$" and "$\tau'$" are frequency independant function, then~\eqref{eq:RelaxFrequency} can be written

\begin{eqnarray}
	\label{eq:RelaxFrequencySigmaDelta}
	\omega_{\tau}(\textup{cm}^{-1})=\frac{N_e}{m^{*}_e}\frac{\rho'}{\delta}\frac{e^2}{2\pi c}
\end{eqnarray}

where $\rho'$ is in $\Omega \cdot \textup{cm}$, $\rho' \neq \rho, N_e$, the electron density (free and valence electrons) and $m^{*}_e$ the effective masses of these electrons in their inertial opposition to the movement.\\
As from reference~\citep{Jackson98_ClassicalED}, cited here as equation~\eqref{eq:Qfactor}

\begin{equation}
  Q = \frac{\mu}{\mu_{conductor}} \left(\frac{V}{S\delta}\right) \times \left(\textup{Geometrical Factor}\right)
  \tag{\ref{eq:Qfactor}}
\end{equation}

Using equations~\eqref{eq:MultimodalCavityResonator} and~\eqref{eq:Qfactor}, and the plasma frequency of the material $X$; \textit{i.e.}

\begin{equation}
	\label{eq:PlasmaFrequency}
	(\omega_{\rho})_X=\frac{1}{2\pi c}\left[\frac{4\pi N_e e^2}{m^{*}_e}\right]^{1/2}
\end{equation}

we can, then, obtain, in substituting~\eqref{eq:PlasmaFrequency} in~\eqref{eq:RelaxFrequencySigmaDelta}, and from~\eqref{eq:Qfactor} and~\eqref{eq:MultimodalCavityResonator}

\begin{equation}
	\label{eq:SubstitutedEquation}
	\sigma^{(1)}_X=\frac{\rho'_X(\omega^2_{\rho})_X}{2}\cdot\frac{S}{V\cdot(geo.fac.)}\cdot\frac{(\mu_c)_X}{\mu}.
\end{equation}

As following~\cite{Jackson98_ClassicalED}, the "geometrical factor" evolutes towards a constant for hight mode number, the term $S/(V\cdot(geo.fac.))$ appears as a constant for a cavity (resonator) of specified form in the analysis for different materials. Relative to gold, equation~\eqref{eq:SubstitutedEquation} implies

\begin{equation}
	\label{eq:GoldSubstituedEquation}
	G_X=\frac{(\omega_{\tau})_X}{(\omega_{\tau})_{Au}}=\frac{\sigma^{(1)}_X}{\sigma^{(1)}_{Au}}=\frac{(\omega^2_{\rho})_X}{(\omega^2_{\rho})_{Au}}\cdot\frac{(\mu_c)_X}{(\mu_c)_{Au}}\cdot\frac{\rho'_X}{\rho'_{Au}}.
\end{equation}

Thus, the $\sigma^{(1)}_X$ must be measured in an equivalent metallic guide in equilibrium propagation and the $\rho'_X$ obtained from~\eqref{eq:GoldSubstituedEquation}. The phenomena of multimodal coupling being related to a strong power exchange from high frequency modes to lower frequency modes, the $\rho'_X$ will, then, correspond to the $\rho_{optical}$ of references~\cite{Ordal83_OptPropMetalsIR,Ordal85_OptProp14MetalsIR}, in the case where $\left(\frac{\mu}{\mu_c}\right)_X\simeq1$. In other cases, as $Fe, Ni, Co$, the result has to be labelled $(\rho'_X, \mu_{c_X})_{optical}$. As for $Au$ and $Ag$, their $\rho_{optical}\simeq\rho_o, (\omega_{\rho})_{Au}\simeq(\omega_{\rho})_{Ag}$ and their $\left(\frac{\mu}{\mu_c}\right)_X\simeq1$,~\eqref{eq:FactorGold} becomes 

\begin{equation}
	\label{eq:FactorGoldReformulation}
	\frac{\sigma^{(1)}_{Ag}}{\sigma^{(1)}_{Au}}\simeq\frac{\rho_{Ag}}{\rho_{Au}}\simeq\frac{(\omega_{\tau})_{Ag}}{(\omega_{\tau})_{Au}}
\end{equation}

Table~\ref{tab:Table 2} shows the results obtained for the normalization of an ensemble of good conductors as based on references~\cite{Ordal83_OptPropMetalsIR} and~\cite{Ordal85_OptProp14MetalsIR}, from where the different measured values of $(\omega_{\tau})_X$, gold assuming a role of reference at a $\textup{T}=293.75 \textup{K}$, \textit{i.e.} $(\omega_{\tau})_{Au}=2.04$cm$^{-1}$. It shows also that relative coefficients $G_X$ are in "accordance" to the ratio $(\omega_{\tau})_X/(\omega_{\tau})_{Au}$ and of $\sigma_X^{(1)}/\sigma_{Au}^{(1)}$.

{
\centering
\begin{sidewaystable}[!htbp]
	\begin{center}
		\begin{tabular}{| >{\centering\arraybackslash}p{2cm} | >{\centering\arraybackslash}p{0.7cm} | >{\centering\arraybackslash}p{1.5cm} | >{\centering\arraybackslash}p{1.5cm} | >{\centering\arraybackslash}p{1.5cm} | >{\centering\arraybackslash}p{2.5cm} | >{\centering\arraybackslash}p{2.5cm} | >{\centering\arraybackslash}p{2cm} | >{\centering\arraybackslash}p{3.5cm} |}
			\hline
			Elements & $\frac{\mu}{\mu_X}\newline\sim 1$ & $\omega_{\tau}^*\cdot 10^{-2}\newline($cm$^{-1})$ & $\omega_p^*\cdot 10^{-4}\newline($cm$^{-1})$ & $\omega_p^2\cdot 10^{-8}\newline($cm$^{-2})$ & $\rho'=\rho_{op}^*\newline(\mu\cdot\Omega\cdot \textup{cm})$ & $\frac{\sigma_X^{(1)}}{\sigma_{Au}^{(1)}}=G_X$ & $\frac{(\omega_{\tau})_X}{(\omega_{\tau})_{Au}}=G_X$ \\

			\hline
			Au & 1 & 2.04$^{**}$ & 7.28 & 53 & 2.32 & 1 & 1 \tabularnewline[10pt]

			\hline
			Ag & 1 & 1.45 & 7.27 & 52.9 & 1.63 & 0.72 & 0.71 \tabularnewline[10pt]

			\hline
			Al & 1 & 6.60 & 11.9 & 141.6 & 2.77 & 3.22 & 3.22 \tabularnewline[10pt]

			\hline
			Cu & 1 & 0.732 & 5.96 & 35 & 1.24 & 0.35 & 0.36 \tabularnewline[10pt]

			\hline
			Pb & 1 & 16.3 & 5.94 & 35.3 & 27.7 & 7.95 & 7.99 \tabularnewline[10pt]

			\hline
			Mo & 1 & 4.12 & 6.02 & 36.2 & 6.82 & 2.01 & 2.02 \tabularnewline[10pt]

			\hline
			Ti & 1 & 3.82 & 2.03 & 4.12 & 55.6 & 1.86 & 1.87 \tabularnewline[10pt]

			\hline
			W & 1 & 4.87 & 5.17 & 26.7 & 10.9 & 2.37 & 2.39 \tabularnewline[10pt]

			\hline
			Pd & 1 & 1.24 & 4.40 & 19.4 & 3.84 & 0.61 & 0.61 \tabularnewline[10pt]

			\hline
			Pt & 1 & 5.58 & 4.15 & 17.2 & 19.4 & 2.71 & 2.74 \tabularnewline[10pt]

			\hline
			
		\end{tabular}
	\end{center}
	* from reference~\cite{Ordal83_OptPropMetalsIR,Ordal85_OptProp14MetalsIR}\\
	** as normalization
	\caption{}
	\label{tab:Table 2}
\end{sidewaystable}
}

\newpage

That analysis shows that a good conductor metallic cavity, when maintained at constant temperature, acts, so far as its spectral density distribution is concerned, as if the field as equilibrium is only "confined" within the volume "$V$", and $\hbar$ being a physical constant related to a normalization in its development in a \textit{complete set of orthonormal functions}. Thus the overlooked finite-spacetime paradigm in blackbody radiation analysis leads to establish the "ad hoc" commutor function of $\hbar$ in quantum physics, in this case, on a sound physical theory, the Maxwell theory. Moreover, it explains the statistical aspect of the vacuum field through multimodal coupling of two kinds, while in the restricted spacetime paradigm of the quantum physics, that aspect rest upon the notion of "quantum vacuum fluctuation". Thus, one cannot assert that the notion of decoherence by multimodal coupling in classical electrodynamics is emerging, in this case, from quantum physics. It is the result of finite-spacetime analysis of electromagnetic fields in multimodal modes and states and the "smearing off" of informations that can be obtained from the systems.

\section{Conclusion and outlook}
\label{sec:Conc&Outlook}
Connections between quantum mechanics and electromagnetism have been established throughout the analysis of the blackbody problem presented in this article. We came to the conclusion that coupling between modes of the cavity, following the coupled power theory for highly multimodal guided propagation, is an essential element in the analysis of the blackbody cavity and sheds new light on the physical interpretation of this century-old problem. The main features of the approach followed were: i) replacing the basis of infinite plane waves by finite TE, TM modes to keep the system of electromagnetic differential equations coupled with space and time, ii) analyzing the interaction of the EM field with the cavity walls based on the temporal lattice of damping frequencies created by the free electron relaxation lifetime, and iii) considering the maximum energy exchange in this interaction occurring at synchronization between electronic damping frequencies and electromagnetic modal frequencies. The Planck constant naturally appears at the outcome of this analysis using material parameters from gold as example, the "path towards equilibrium" providing a classical value for $\hbar$. We have also shown that the same process applies to a large group of good conductors metallic cavities. It follows that the Planck constant can now be derived as a normalization factor between thermal and electromagnetic energy exchanges at thermal equilibrium and the "ultraviolet catastrophe" is no longer an unexplained phenomenon in classical electrodynamics and it shines new light on the notion of decoherence. We provide here only a brief outlook on some other basic issues.

\subsection{Fluctuations}
Our focus here was on the mechanism proposed to reach equilibrium in the cavity and the Boltzmann distribution has been the only result used from statistical physics. We however want to emphasize that statistical mechanics is still fully applicable to the blackbody cavity problem governed by coupled power theory. This theory encompasses power fluctuations for an electromagnetic mode, or a subset of a few modes, away from the average power distribution of the entire set, hence preserving the stochastic nature of the system. Additionally, mode coupling enables a fast redistribution of power between them, hence accelerating the evolution of the system toward equilibrium. As stated previously, if the initial power distribution is unknown, it is not possible to apply a reversibility principle to such systems.

\subsection{Uncertainty principle}
\label{subsec:Uncert}
Observable quantities of TE-TM waves as derived here from the coupled power theory and experimental measurements will obey the general recognized uncertainty properties for all EM waves:
\begin{eqnarray}
	\label{eq:uncertainty}
  \Delta \nu \Delta t & \geqslant & \frac{1}{4\pi} \; ,\\
  \Delta x \Delta k & \geqslant & \frac{1}{4\pi} \; .
\end{eqnarray}
Relation~\eqref{eq:uncertainty} is readily seen to correspond to Heisenberg's uncertainty principle $\Delta E \Delta t \geqslant \hbar /2$ since we also derived $E=h \nu$ as the quantization of thermal and electromagnetic energy exchanges. This suggests to interpret the uncertainty principle as a consequence of the finite space-time for the EM waves: without the infinite plane wave, there is no more pure frequency or wavelength, each finite wave excitation is associated to a frequency bandwidth and finite extent in space. It is possible to go beyond this first order approximation by taking into account the group velocity $v_{gr}$ of TE-TM waves in the the $\Delta x \Delta k$ inequality through the mode momentum $p=E/v_{gr}= h\nu /c \sqrt{1- \omega_{c}^2 /\omega^2}$ where $\omega_c$ is the cutoff frequency. 

\subsection{Photoelectric effect}

The photoelectric effect has traditionally been one of main justifications for the photon hypothesis to represent electromagnetic radiation. However in more recent years, it was shown that a semi-classical theory where only the photoelectric detector is quantized in terms of $\hbar$ embodies all the features of the photoelectric effect~\cite{Loudon00_QTheoryLight}. In this article, we presented further evidence that the appearance of $\hbar$ in light absorption, emission and coupling does not call for the introduction of the photon concept. Also, temperature dependence of the photoelectric threshold is more directly linked to material properties in our analysis by using the damping frequency. Finally, we emphasize that the dominant physical mechanism in our approach is ultra fast coupling compared to slower absorption and re-emission, hence matching characteristic time scales of the photoelectric effect.

\section{Acknowledgments}
This work was supported by the Natural Sciences and Engineering Research Council of Canada (NSERC). We also acknowledge the precious help of Gabrielle Th\'{e}riault, Alexandre April, Jean-Rapha\"el Carrier, Pierre Tremblay and the recent collaboration of Anthony Richard.

\bibliography{Tremblay9juin}

\end{document}